\documentclass[10pt,letterpaper]{article}
\usepackage[top=0.85in,left=0.75in,footskip=0.75in]{geometry}

\usepackage{amsmath,amssymb}

\usepackage{changepage}

\usepackage[utf8x]{inputenc}

\usepackage{textcomp,marvosym}

\usepackage{cite}

\usepackage{nameref,hyperref}

\usepackage{microtype}
\DisableLigatures[f]{encoding = *, family = * }

\usepackage[table]{xcolor}

\usepackage{array}
\usepackage{caption}
\usepackage{subcaption}
\usepackage[numbers]{natbib}

\newcolumntype{+}{!{\vrule width 2pt}}

\newlength\savedwidth

\raggedright
\usepackage[aboveskip=1pt,labelfont=bf,labelsep=period,justification=raggedright,singlelinecheck=off]{caption}

\makeatletter
\renewcommand{\@biblabel}[1]{\quad#1.}
\makeatother

\date{}

\usepackage{lastpage,fancyhdr,graphicx}
\usepackage{epstopdf}

\fancyheadoffset[L]{2.25in}
\fancyfootoffset[L]{2.25in}

\begin{document}
\vspace*{0.2in}

\begin{flushleft}
{\Large
\textbf\newline{Dynamics of Investor Spanning Trees Around Dot-Com Bubble} 
}
\newline
\\
Sindhuja Ranganathan\textsuperscript{1,*},
Mikko Kivel\"a\textsuperscript{2},
Juho Kanniainen\textsuperscript{1}
\\
\bigskip

\textbf{1} Industrial and Information Management/Tampere University of Technology, Tampere, Finland
\\
\textbf{2} Department of Computer Science, School of Science/Aalto University, Espoo, Finland
\\

\bigskip

%
%


Current Address: Industrial and Information Management/Tampere University of Technology, Tampere, Finland 



* sindhuja.ranganathan@tut.fi

\end{flushleft}
\section*{Abstract}
We identify temporal investor networks for Nokia stock by constructing networks from correlations between investor-specific net-volumes and analyze changes in the networks around dot-com bubble. 
We conduct the analysis separately for households, non-financial institutions, and financial institutions. Our results indicate that spanning tree measures for households reflected the boom and crisis: the maximum spanning tree measures had clear upward tendency in the bull markets when the bubble was building up, and, even more importantly, the minimum spanning tree measures pre-reacted the burst of bubble. At the same time, we find less clear reactions in minimal and maximal spanning trees of non-financial and financial institutions around the bubble, which suggest that household investors can have a greater herding tendency around bubbles.



\section*{Introduction}
The strategic interaction and collection of individuals or agents in a financial setup can play a key role in determining their financial outcomes. Understanding how investors behave and operate has been a topic of interest in behavioral finance in the recent past. Earlier in the literature investors trading strategies and investor behavior were studied at an aggregated level using conventional  regression methodologies \cite{grinblatt2000investment,odean1998investors,brennan1997international,kaniel2008individual,barrot2016retail,hoffmann2013individual,chiang2010empirical}. The evolution of networks of stocks and currency rates and their structural change have been successfully analyzed in the existing literature\cite{mantegna1999hierarchical,onnela2003dynamic,naylor2007topology,heimo2009maximal,emmert2010influence,song2011evolution}. Effect of economic and financial bubble on the stock market have been analyzed in the literature \cite{zhou2009case, zhou20032000,jiang2010bubble}. However, {\em investor networks} have been examined much less, 
and even though complex network methods have been applied to identify investor networks recently\cite{ozsoylev2014investor,tumminello2012identification}, we still lack research to study the dynamics of investors network around a financial crisis.
This paper aims to be the first step to reveal understanding on investor networks by focusing on the dynamics of correlation networks of investors over the Dot-com (IT Millennium) bubble using unique investors transaction registry data which contains all the trades of Finnish households and institutions in Helsinki Exchange. We especially focus on the question of how gradual and non-gradual changes in investor network structure are related to the stock price process. This research opens avenues to reveal understanding on actual mechanisms of stock markets to identify domino effects that can propagate through investors and propels the stock markets into a crisis state.

In this paper investors correlation matrix is obtained using time series of investor-specific daily net volumes for Nokia, one of the most important technology company around the millennium. At the same time, Nokia is the most liquid stock in our data sample from Helsinki stock exchange and there has been other research based on Nokia's stock market data, for example in refs. \citen{tumminello2012identification,kalev,lillo2015news}. Investors' correlation matrices are estimated for three main categories of investors: financial institutions, households and non-financial institutions. 
Correlation matrices can be interpreted as link-weighted networks and the links in the resulting networks where all nodes are connected can be filtered with a multitude of different approaches \cite{Tumminello2005Tool,Serrano2009Extracting,Chi2010Network,emmert2010influence}. An elegant, and popular method in stock market network analysis, is to employ minimal or maximal spanning tree methods to find a ``backbone'' of the full correlation network \cite{mantegna1999hierarchical,Vandewalle2001Nonrandom,onnela2003dynamic,heimo2009maximal,Wang2016Correlation,Birch2016Analysis}. Several more complicated correlation matrix construction and filtering methods have been developed more recently \cite{Tumminello2005Tool,Chi2010Network,Kenett2012Dependency,Xi2015Detrended,Nakajima2015Dynamic,Musmeci2016Multiplex,Kwapien2017Minimum}, but utilizing these is left for the future research.


The analysis of dynamics of investor networks developed in this paper introduces two theoretical challenges when compared to other financial correlation networks.
First, the set of investors is much larger than, for example, the number of stocks, and the set of active investors is strongly time-varying. The vast majority of methods developed for analyzing dynamic, or temporal, networks are based on the assumption that only the links change and the set of nodes are stable \cite{Holme2012Temporal,Holme2015Modern}. Further, changes in the set of investors even limits the applicability of methods based on analyzing each network snapshot separately, because metrics that are sensitive to network size cannot be compared across different time windows where the number of investors can be exceedingly different.
The second challenge is related to the widely varying sparsity of the time series where few investors are extremely active and many others trade very infrequently. 
The active investors could be investigated using a high temporal resolution and short observation window lengths, but the infrequent investors need a lower resolution and a longer time window. 
The conventional correlation analysis done here requires a single time resolution level and observation window length to be chosen, and this choice must be a compromise between the two extremes.

We construct minimum and maximum spanning trees for networks within six month time windows with displacement of one month. 
Our results with estimated correlations between households' transactions show that the average weight of maximum spanning tree increases and average weight of minimum spanning tree decreases before the tipping point of the stock prices (at which stock prices start to decline), after which they remain quite stable. 
In other words, when the bubble propagates, then, on average, an investor has a more and more positive correlations with another investors in the maximum spanning tree, but, at the same time, the correlations with the most distant investor, in terms of trading style, becomes even more negative in the minimum spanning tree. 
This suggests that households became polarized before the Nokia prices crash in 2000. However, as strong effect cannot be observed for financial institutions -- the average weights of minimum and maximum spanning trees of institutional investors are not as clearly related to the evolution of financial crisis.

\subsection*{Dot-com Bubble}
In this paper, we analyze the behavior of Nokia's investors around the dot-com bubble of year 2000. Bubbles are phenomenon when price of assets deviate from their fundamental values \cite{kindleberger1991bubbles}. Generally, during bubbles, investors purchase shares anticipating future gains and when bubbles collapse it leads to sudden fall in the prices, which was the case also in dot-com bubble. 
Particularly, during the late 1990s, internet-based stocks dominated the equity markets and there were lots of investments in the internet and technology based start-ups with extremely optimistic expectations. As people started pouring money on technology based start up companies, price of their share in the stock market grew very high. During early 2000, investments in these companies reduced drastically and many of these companies that were expected to generate profits failed, leading to the bubble to burst. As a consequence of this, there was panic selling and market got slumped. 

Bubbles have been studied quite extensively and from various perspectives in the literature. According to ref. \citen{johansen1999log}, market prices during bubbles follow power-law acceleration and have log-periodic oscillations. Dot-com bubble had similar characteristics and ended up in crash (see, for example, ref. \cite{johansen2000nasdaq}).  One perspective is that bubbles occur due to the uncertainty that prevail in the market. \cite{oechssler2011ingredients} 
In this regard, ref. \citen{pastor2006there} provides evidence that uncertainty is plausible for a sudden rise in the price of some stocks as high level of uncertainty matched high prices and high return volatility in the market during dot-com bubble. The sudden rise and fall in the market prices during dot-com bubble was associated with variations in risks from various sources. Bakshi and Wu \cite{bakshi2010behavior} show that with the rising valuation of the NASDAQ 100, return volatility as a risk measure increased, the estimates for the market price for diffusion risk became negative (from September 21, 1999, to January 5, 2000), and the market price of jump risk became unusually high. Another perspective of bubble's occurrence is that it occurs when there are new innovations \cite{perez2009double} that investors see as opportunity pulls, expecting high profits in the future. Other reasons for the occurrence of bubble are lack of experience in traders  \cite{dufwenberg2005bubbles}, investor's emotions \cite{andrade2015bubbling}, investor's over-confidence \cite{abreu2003bubbles} and public announcements \cite{corgnet2010effect}. There are several reasons for a bubble to burst. According to ref. \citen{pastor2006there}, one of the reason for the dot-com bubble to burst was that the expected profitability of technology stocks became low. Not all bubbles leads to crashes, but when a bubble crash, it signals important information to the market. According to ref. \citen{perez2009double}, when a bubble bursts it signals that there is a need to implement new innovations that happened in bubble period. This requires  social and economic support  to continue the growth of innovations which could benefit the economy.

\section*{Results}


Next we describe how we construct a series of correlation networks of investors investing in Nokia stock around the Dot-Com Bubble, 1998 - 2002, and report the basic statistics related to the changes in these networks. We then continue to investigate the minimal and maximal spanning trees we extract from these fully connected networks. We report the results of our analysis separately for Finnish households, financial institutions and non-financial institutions.


\subsection*{Nodes in the networks: Active investors}

\begin{figure}[ht!]
\begin{center}
\begin{subfigure}{0.5\textwidth}
\includegraphics[width=1\textwidth]{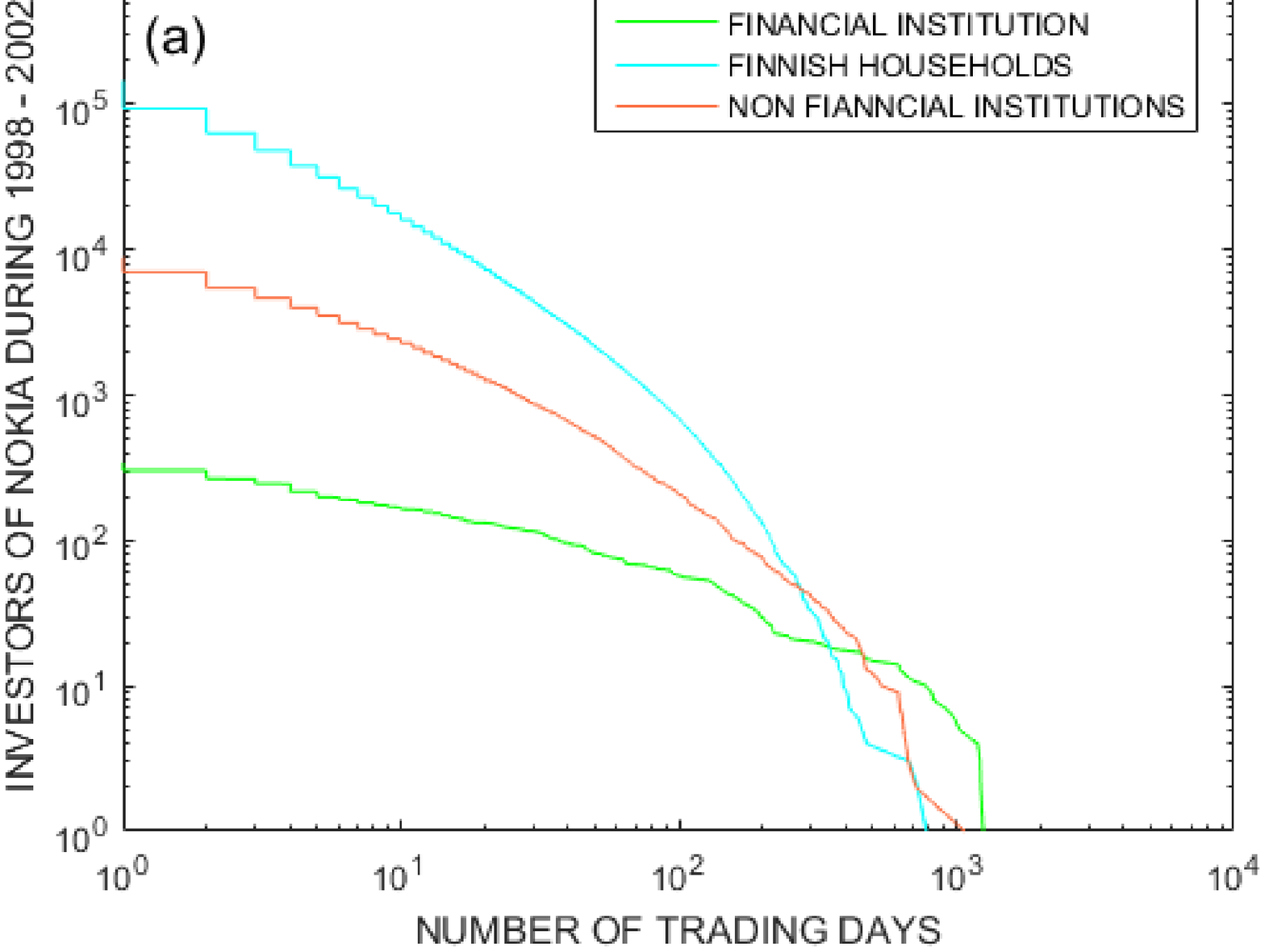}
\captionsetup{labelformat=empty} 
\label{fig:no_of_nodes_2000_a}
\vspace{-5\baselineskip}
\end{subfigure}


\begin{subfigure}{0.45\textwidth}
\includegraphics[width=1\textwidth]{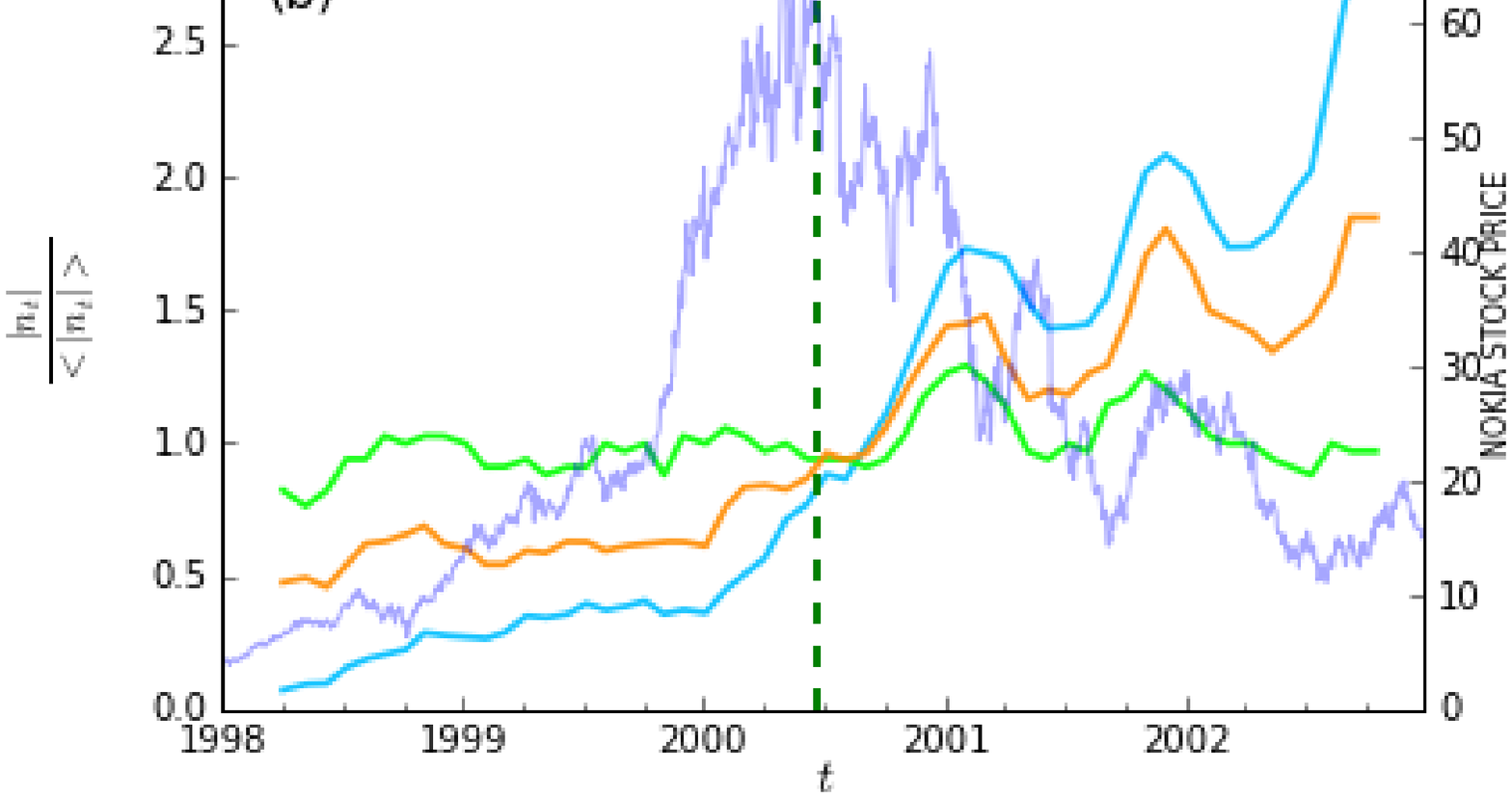}\captionsetup{labelformat=empty}  \label{fig:no_of_nodes_2000 b}
\vspace{-12\baselineskip}
\end{subfigure}
\begin{subfigure}{0.45\textwidth}
\includegraphics[width=1\textwidth]{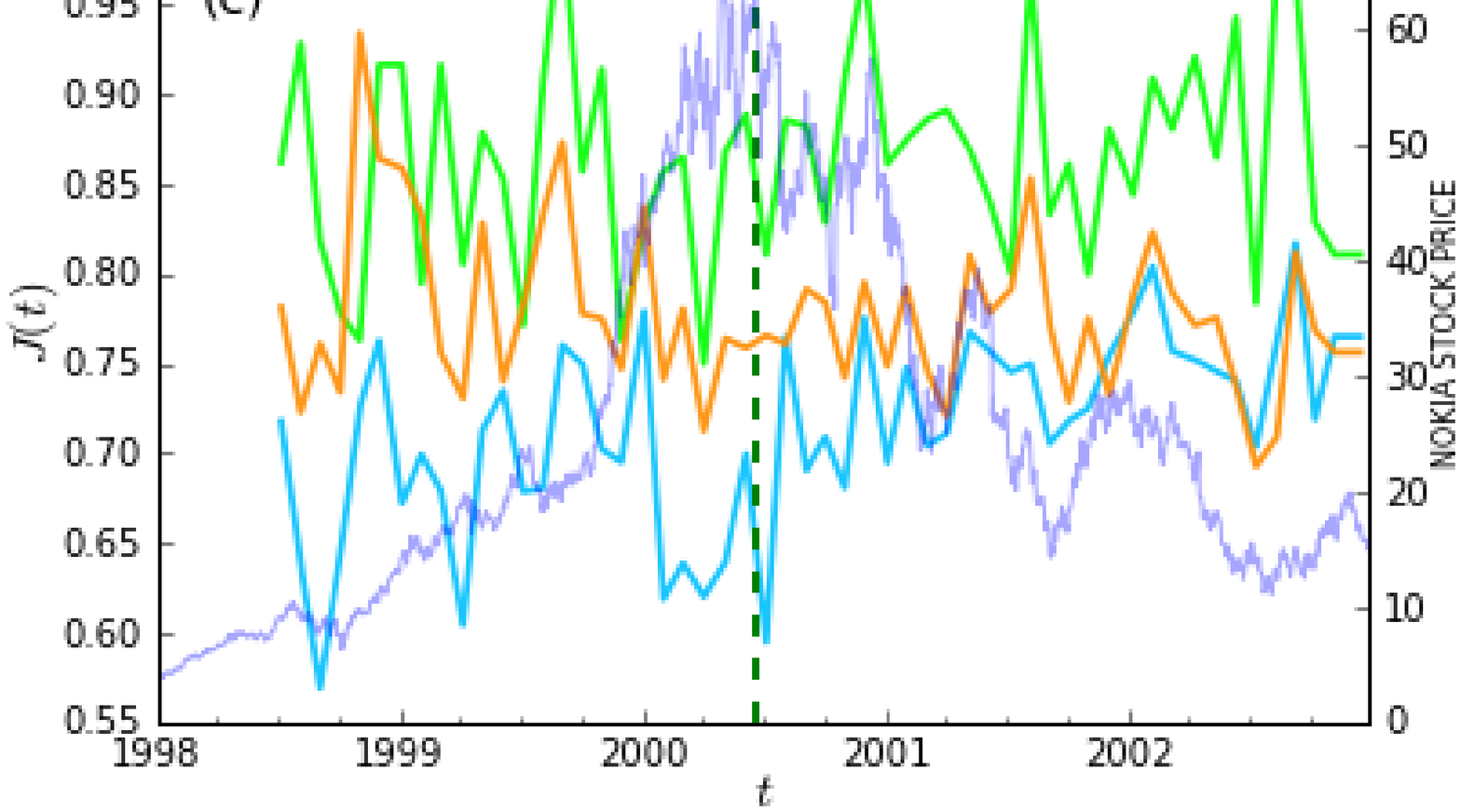}\captionsetup{labelformat=empty} 
\label{fig:no_of_nodes_2000 c}
\vspace{-12\baselineskip}

\end{subfigure}

\caption{ 
The number of investors in Nokia stocks during full time period 1998 -- 2002 and the change of investors across the 6 month time windows. 
(a) Cumulative distributions of investors and their respective trading days during the full time period. 
 (b) The evolution of number of investors trading Nokia in the 6 month time windows for households, non-financial institutions, and financial institutions. 
The numbers of investors in each category $|n_t|$ are very different across categories, and they are normalized by the average numbers of investors in the full time period $\langle |n_t| \rangle$.
(c) The change of investors measured using Jaccard coefficient $ J(t) = \frac{|{n_{t+1} \cap \, n_{t}|}}{|n_{t+1} \cup \, n_{t}|} $, where $n_{t}$ and $n_{t+1}$ represent the sets of nodes in the network of months $t$ and $t+1$, respectively, for different investor categories and 6 month time windows. 
The value of $J(t)$  is higher (lower) the more (less) similar the consecutive networks are. 
Results for each time window in panels (b) and (c) are plotted at the end of the window. That is, each point is estimated with data over the previous 126 trading days (6 months). The estimation windows are rolling by one month, and the resulting points are joined by solid lines.
In panels (b) and (c) the green dotted vertical line in the figures represents the highest stock price of Nokia in the sample period, and the blue curves (with axis on the right) represent the Nokia stock price.  
In all panels, lime-green curve corresponds to financial institutions, cyan curve to households and orange curve to non-financial institutions.
}
\label{fig:no_of_nodes_2000}
\end{center}
\end{figure}

The nodes in the networks we construct are investors, and in order to estimate the correlations between pairs of them we need to have enough data on their trading behavior.
Figure \ref{fig:no_of_nodes_2000}a depicts the distribution of investors in the period 1998--2002 and shows that there are many investors who have traded only for few days but relatively few investors who have traded for many days, making the data sparse. We take two steps to alleviate the problems related to sparse data in the network construction: First, we only consider \emph{active investors} who have traded minimum of 20 days in a given time period.
 Second, daily net-volumes of each active investor is averaged over a week (that is, we apply investor-specific simple moving averages).

We investigate our total sample period 1998--2002 by moving a 6 month sliding time window on it. 
By using the above definition for active investors for each 6 month time window, Figure \ref{fig:no_of_nodes_2000}b depicts the evolution of the number of active financial institutions, households and non-financial institutions on these time windows. 
We see that the numbers of active households and non-financial institutions had positive trends over the sample period while the number of active financial institutions remained rather stable. Importantly, the bubble ``burst'' did not have clear effects on the number of active traders. 

Even though the number of investors in each time window can be stable, the set of investors can vary significantly. This is indeed the case, as shown by
the Figure \ref{fig:no_of_nodes_2000}c where we use the Jaccard index to investigate the number of investors overlapping in the every subsequent time window.
Note that the activeness criterion (at least 20 observation in six months) is applied for each estimation period with a displacement of one month, and that this filtering has an effect to the Jaccard index values.
We observe that the networks of households have lower similarity between each other compared to financial institutions, meaning that the turnover of active household investors is relatively high over time. 
This means that, especially for relatively inactive household investors, the networks in different time windows are bound to be very different, and if we observe any stable in network statics they cannot be only explained by stability of the networks, but they need to be explained by some other organizing principles in the system.

\subsection*{Links in the networks: Correlations in trading patterns}

We use the Pearson correlation of trading patters of investor pairs inside each time window to construct a links between the investors (for details, see Methods).
The Pearson correlation coefficient has been used extensively in the network analysis of time series of stock prices \cite{mantegna1999hierarchical} and it has some clear advantages also in the analysis individual investor trading.
Observations with exceptionally high trading volumes can represent days with arrival of important information, which are of our interest to analyze if investors react to the information in the same way, and therefore it is a desired property that the measure is sensitive to exceptionally large values. 
In contrast to Pearson correlation, Kendall and Spearman correlations are looking at rank-order as opposed to metric information, and thus they do not weight these outlier days appropriately. 

\begin{figure}[ht!]
\begin{center}
\minipage{0.45\textwidth}
\includegraphics[width=\textwidth, angle=0]{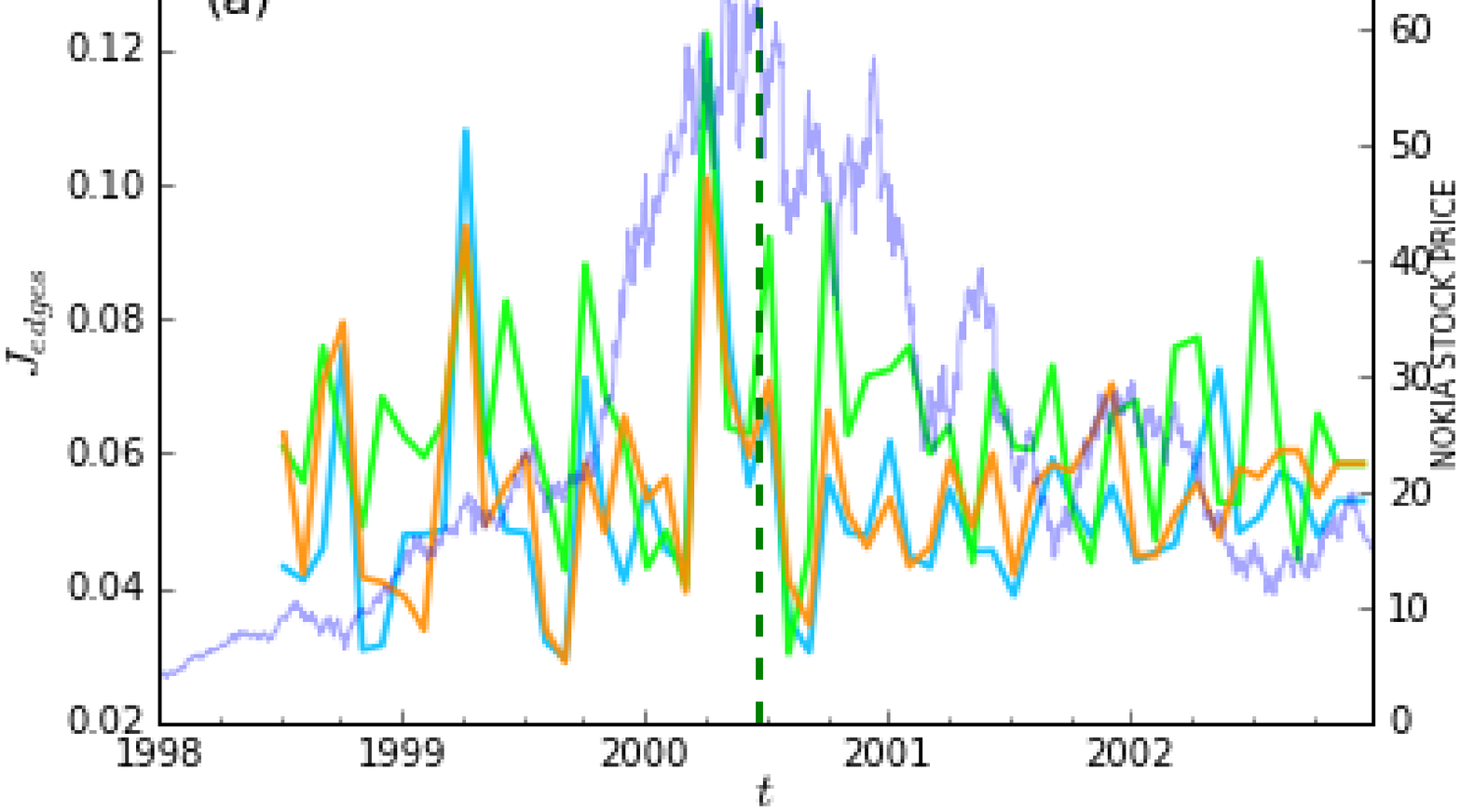}
\captionsetup{labelformat=empty} \label{jacc_chg a}
\vspace{-10\baselineskip}
\endminipage
\minipage{0.45\textwidth}
\includegraphics[width=\textwidth, angle=0]{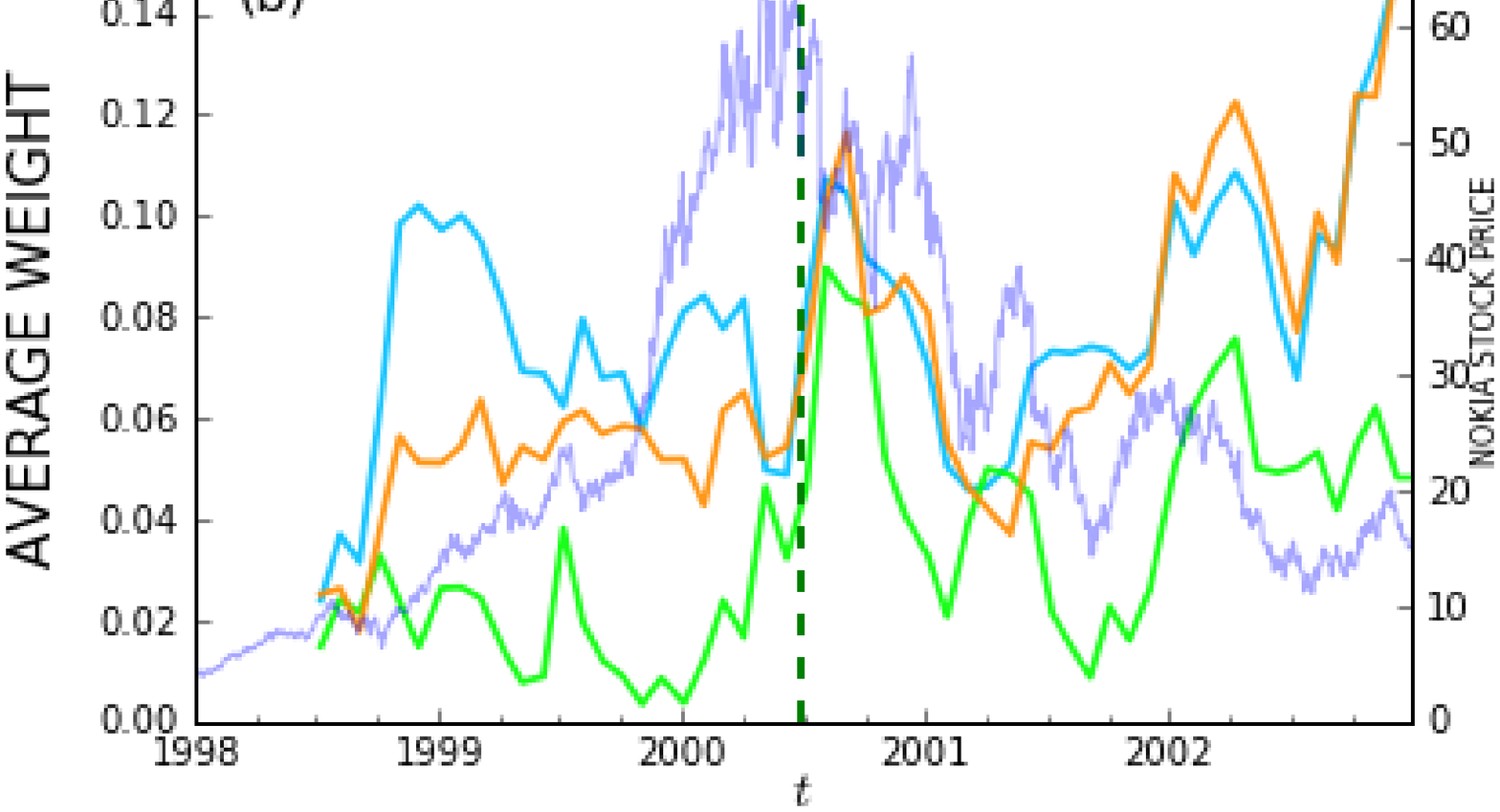}
\captionsetup{labelformat=empty}
 \label{jacc_chg b}
\vspace{-10\baselineskip}
\endminipage
\end{center}
\caption{
The change in investor correlations of Nokia stock trading across the 6 month time windows during 1998--2002. 
(a) The average change in correlations between two consecutive time windows $ J_{edges}(t)$ (see Eq. \ref{eq:jaccard_edges} in the Methods section).
(b) The average edge weight, or correlation, in each time window. 
Every point is estimated with data over the previous 126 trading days (6 months), and the estimation windows are rolling by one month. The green dotted vertical line represents the highest stock price of Nokia in the sample period, and the blue curves (with axis on the right) represent the Nokia stock price. 
The lime-green curves correspond to financial institutions, cyan curves to households and orange curves to non-financial institutions.
} 
\label{fig:jacc}
\end{figure}

Not only the nodes change between the different time windows, but also the weights of the links (the correlations) are relatively unstable. To quantify this, we measure the average absolute change in correlations between nodes that remain in two consecutive time windows (see Eq. \ref{eq:jaccard_edges} in Methods) and the average correlation between all pairs of nodes in Figure \ref{fig:jacc}.
The figure \ref{fig:jacc}a demonstrates the average change in the correlations among pairs of investors who both appear in subsequent periods. 
The change in correlations between two consecutive time windows is of the same order as the standard deviation of the correlations inside the time windows. That is, the network is relatively unstable in its links, but, as we will see in the next, the global organization of the network and related statistics are still rather stable.


\subsection*{Minimum/maximum spanning trees}

The correlation matrices of investors' net trading volumes which we produce can be interpreted as weighed networks where all node pairs (i.e. investors) are connected. Particularly, investors $i$ and $j$ are connected by a weight of $\rho_{ij} \in [-1,1]$, which is the Pearson correlation coefficient between investors' daily netvolumes. Clearly, the topological structure of these fully connected graphs is trivial and all the information is in the weights. In order to analyze the structure, one needs to filter out parts of the edges, and various approaches for doing that exist \cite{Tumminello2005Tool,Serrano2009Extracting,Chi2010Network,Kwapien2017Minimum}. Following the literature of analysis of stock prices \cite{mantegna1999hierarchical,onnela2003dynamic} we employ one of the simplest filtering methods and construct maximum (and minimum) spanning trees of correlation networks.

The idea of maximum spanning tree analysis is to filter out as much edges as possible such that the network is still connected and the highest possible weights (or, correlations) are not filtered out (for details, see the Methods section). According to Ref. \citen{ozsoylev2014investor}, information links may be identified from realized trades and thus  traders identified with similar trading behavior can have an (private) information channel. In the light of this idea about the inference of information transfer in investor networks, maximum spanning tree would reflect the smallest set of interactions which connect all investors and still have the strongest information flow between them. The interpretation of the empirical investor network as the information network, however, can be questioned as two investors can certainly trade in the same directions without even knowing each other, if they just follow the same investment strategies with the same public information channels. Generally speaking, maximum spanning tree picks the most similar trading strategies while keeping the graph connected, reflects it actual information channels or not. The average weight of maximum spanning tree shows how investors, on average, are pulled together or dispersed in a connected graph, and this quantity has been previously shown to react to crisis in stock price correlations \cite{onnela2003dynamic}. The minimum spanning tree, on the other hand, reflects distant trading strategies, and the average weight of minimum spanning tree can be used to analyze divergent trading strategies in a connected graph of investors.

\begin{figure}[ht!]
\begin{center}
\includegraphics[width=0.9\textwidth, angle=0]{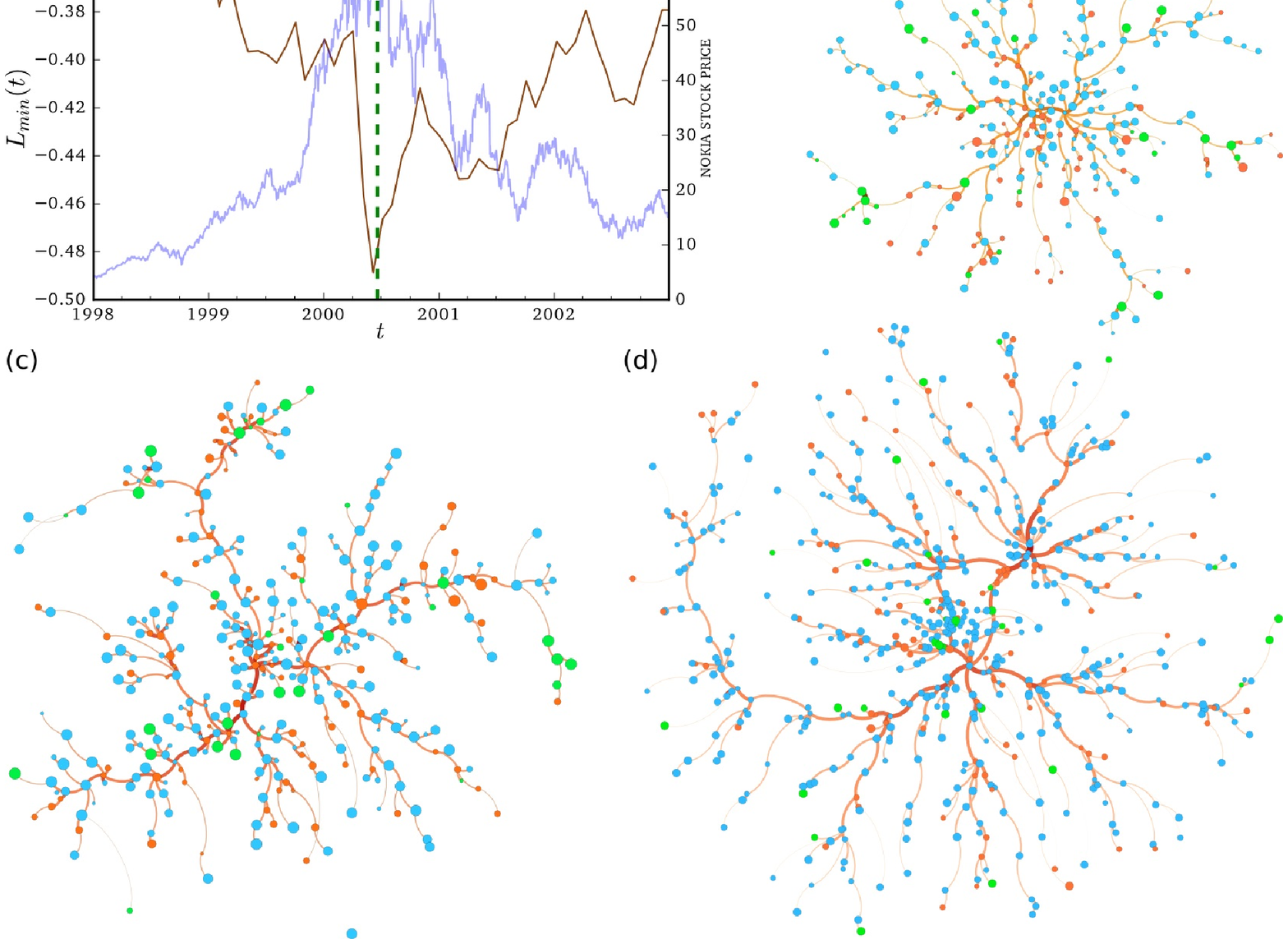}\vspace{-20\baselineskip}
\caption{
The minimum and maximum spanning trees of all investors.
(a) Backward looking average weight of minimum spanning tree, $L_{min}(t)$, for the merged set of investors with 6 month time windows during  1998 - 2002 (brown line).  Every data point is estimated with data over the previous 126 trading days (6 months), and the estimation windows are rolling by one month. The green dotted vertical line in the figures represents the highest stock price of Nokia in the sample period, and the blue curves (with axis on the right) represent the Nokia stock price.
Maximum spanning trees between 
(b) 8-July-1999 and 04-January-2000 (before the crisis), (c) 5-January-2000 and 06-July-2000 (during the crisis), and 
(d) 7-July-2000 and 04-January-2001 (after the crisis). Cyan nodes represents households, orange nodes non-financial institutions, and lime-green nodes financial institutions. Size of the nodes are based on the volume traded by the investor during the period. However, one should not compare the sizes of nodes between different network as the sizes are not comparable across panels. 
 }\label{fig:L_min_all}
\vspace{-10\baselineskip}
\end{center}
\end{figure}

Figure \ref{fig:L_min_all}a shows the evolution of the average weight of minimum spanning tree, $L_{min}$, for the {\em merged} network of investors of the three categories. 
There is an obvious, downward jump in $L_{min}$ just before the tipping point, which is defined as the highest price of the stock Nokia during the sample period.
Importantly, $L_{min}$ is estimated using data from the past, and therefore, no information about forthcoming bubble burst was used. 
That is, the investors pre-reacted to the impending decline in the stock price, and next we focus on investigating which investor groups are behind this reaction.
We visualize the maximum spanning trees in \ref{fig:L_min_all}b-d.
There does not seem to be any clear visible clustering of the categories similar to business sectors in stock networks or geographical regions in currency networks \cite{mantegna1999hierarchical,Heimo2008Detecting,wang2013statistical}.
However, we can see that there might be some local tendency for nodes from the same category to be adjacent, but this observation is not investigated further here.

\begin{figure}[ht]
\begin{center}
\minipage{0.45\textwidth}
\includegraphics[width=\textwidth, angle=0]{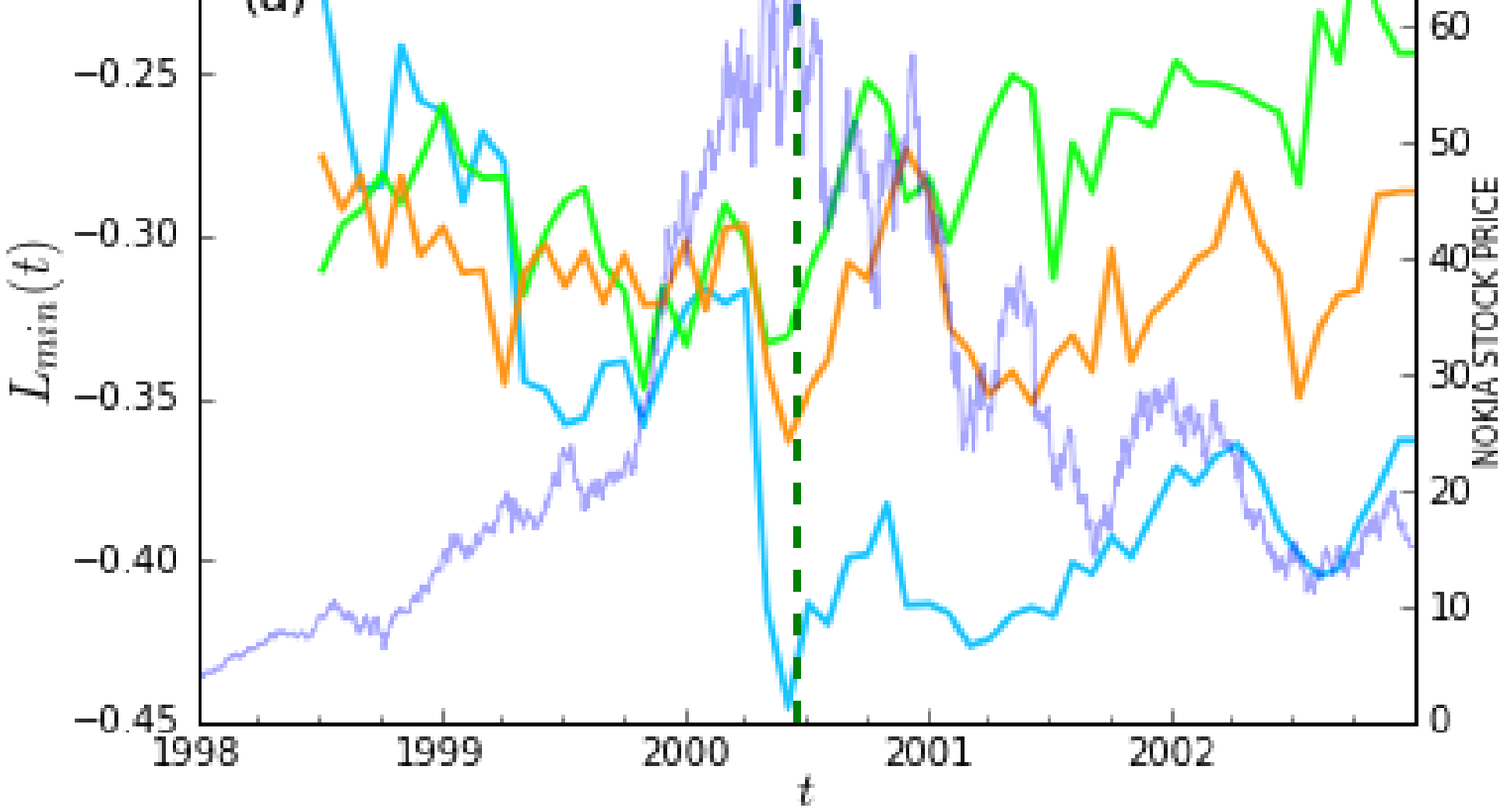}
\captionsetup{labelformat=empty} \label{back_nor_len_2000 a}
\endminipage
\minipage{0.45\textwidth}
\includegraphics[width=\textwidth, angle=0]{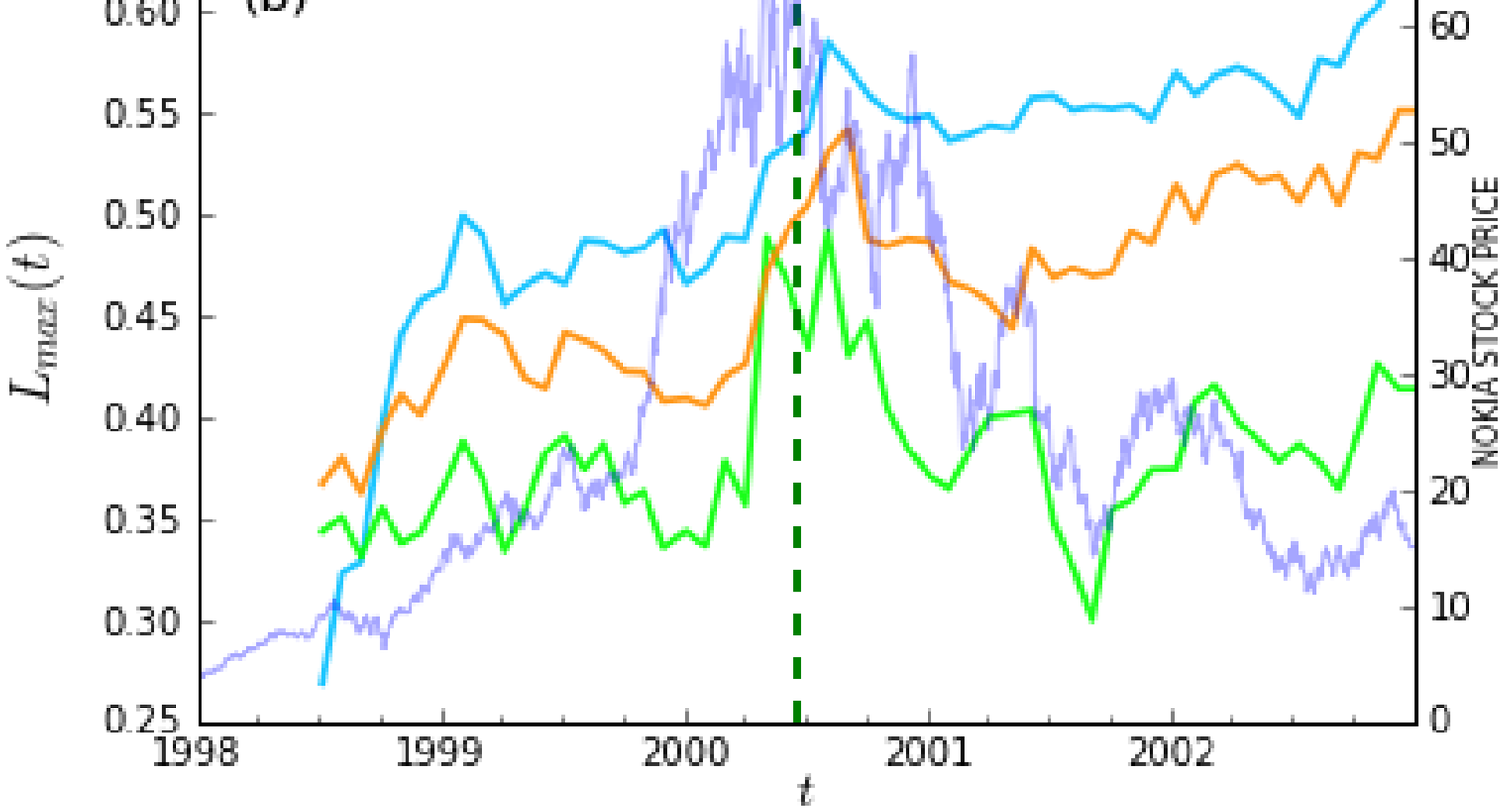}
\captionsetup{labelformat=empty}
 \label{back_nor_len_2000 b}
\endminipage
\caption{Backward looking average weight of the (a) minimum spanning tree, $L_{min}(t)$,  (b) maximum spanning tree, $L_{max}(t)$ for different investor categories with 6 month time window during  1998 - 2002. Every data point is estimated with data over the previous 126 trading days (6 months), and the estimation windows are rolling by one month. The green dotted vertical line in the figures represents the highest stock price of Nokia in the sample period. The lime-green curve corresponds to the plot for Finnish financial institutions, cyan curve corresponds to the plot for Finnish households and orange curve corresponds to  the plot for Finnish non-financial institutions. 
}
\label{back_nor_len_2000}
\end{center}
\end{figure}

Figure \ref{back_nor_len_2000} displays the average weights of minimum and maximum spanning trees, $L_{min}$ and $L_{max}$, around the crisis for networks containg nodes only from one of the three investor categories. Again, every data point is estimated with data over the previous 126 trading days (6 months), and the estimation windows are rolling by one month. Figure \ref{back_nor_len_2000}a shows  that  the average weight of minimum spanning tree, $L_{min}$, of household network suddenly jumps down just some months prior the turning point of the stock price evolution around the crisis. Particularly, the value of $L_{min}$ was -0.32 on 03-April 2000 whereas it was -0.45 on 06-June 2000, after which the stock prices started to burst. Importantly, the difference is considerably large in comparison to other changes in the data sample, yet the estimates, -0.32  and -0.45, are based on partially overlapping estimation data (the length of the estimation period is six months and the analysis is ran with a rolling window of 1 month). Another important observation is that the level of $L_{min}$ does not recover back to the level it was prior to the tipping point during the following two years.  For non-financial and financial institutions, we see no obvious patterns in $L_{min}$ around the crisis. Overall, weights in minimum spanning tree among households are, on average, abnormally negative just around the turning point for households. This means that households, on average, have neighbors in the minimum spanning tree who are trading in an abnormally opposite way. 

Dynamics of maximum spanning trees in Figure \ref{back_nor_len_2000}b provide a slightly different story compared to minimum spanning tree dynamics. Particularly, we see that the average weight of maximum spanning tree, $L_{max}$, for households has a clearly positive trend  prior the spike of February 2000, after which it remains quite stable. Particularly, its value was 0.27 in 1998, it increased almost to 0.6 in two years in bull markets, which is an increase of 122\%! This means that there are investors that have been coming together when the bubble was building up. A positive pre-trend and rather stable post-trend can also be identified for non-financial institutions, but it is weaker compared to households. Financial institutions, however, behave differently regarding $L_{max}$ -- there is a peak in $L_{max}$ for financial institutions just before the tipping point, which lasts a half year, but otherwise $L_{max}$ is relatively stable over the period.
Note that the average weights of the networks displayed in Figure \ref{fig:jacc}b do not display peaks at same times or of same magnitude.

In the light of private information channels that investors use in trading in stock markets (see ref. \citen{ozsoylev2014investor}), our results from maximum spanning tree analysis would suggest that especially household investors' connections to the most important neighbors in a connected graph became more and more important when the techno bubble was building up, which can indicate herding in stock markets. Also the existing literature provides evidence that spanning trees for different financial networks react around financial crises, though with different data sets (and thus with different networks) compared to the present research (see refs. \citen{onnela2003dynamic,coelho2007evolution} with the data on stock returns, \citen{song2011evolution} with data on stock market indexes, and \citen{jang2011currency} with the data on currency exchange rates).

\section*{Discussion}
This paper examines the behavior of Finnish investors using of shareholding registration records for Nokia stock in Helsinki stock exchange from year 1998 to year 2002, which includes the period of the dot-com bubble.
 Analysis for households, non-financial institutions, and financial institutions are conducted using minimum and maximum spanning trees constructed from correlations between investor-specific net-volumes. 
We find that the spanning tree measures reflected the bubble with the data for households, and, in fact, they pre-reacted on forthcoming bear markets, while non-financial and financial institutions show no equally clear reactions. Particularly, the average correlations of households' minimum spanning tree clearly jumped down a couple of months before the Nokia price started to have a negative trend. On the other hand, the average correlation in households' maximum spanning tree dynamics did not jump suddenly right before the burst of the bubble -- rather, the average correlation had a considerably large upward trend in bull markets, increasing from 0.27 to almost 0.60 during two years before the stock price crash, after which it stayed quite stable. This result on maximum spanning trees can reflect information channels between individual household investors -- investors' connections to the most important neighbors in a connected graph became more and more important when the techno bubble was building up, which can indicate herding in stock markets, especially among household investors. 

There are some restrictions in our research on correlated investors network, which are mainly related to how the networks are constructed. We used data on investors' transactions with only one stock, because the other stocks in the our data set are too illiquid to have enough data estimating investor-specific networks. In the future studies multiple similar stocks could be pooled together or methods that function better under sparse data could be used.
Another limitation is the way we used Pearson correlation between the investment time series to calculate the similarities between nodes. There are more sophisticated ways of inferring the latent relationships between the nodes in the literature \cite{Kenett2012Dependency,Xi2015Detrended,Nakajima2015Dynamic,Musmeci2016Multiplex}, but the particular difficulty in the investor networks is the high variations in the transaction frequencies between investors. The high frequency nodes can be analysed with much higher temporal resolution than the low frequency ones, and choosing a single resolution level is a compromise between these two extremes. Finally, the spanning tree analysis discards valuable data in very aggressive way in order to make the system less complex, and there are multiple alternatives in the literature where more data is kept \cite{Tumminello2005Tool,Serrano2009Extracting,Chi2010Network,Kwapien2017Minimum}.
In the future research, we aim to build the network in more sophisticated way, which allows us to analyze a large number of stocks with alternative methods. 


The network of investors is dynamically changing, and the approach taken here---which is in line with the literature on stock correlation networks---was to calculate various static network metrics on snapshots of the network, and then inspect how these metrics change in time.  Methods that do not rely on static networks but measure the dynamics of networks have been developed in the field of temporal networks \cite{Holme2012Temporal,Holme2015Modern}, but most of these approaches have been constructed for networks where the links change dynamically but the nodes are relatively stable. There are, of course, other systems with long temporal data and large changes in the set of nodes, such as citation networks and collaboration networks \cite{Travis2013Coauthorship,Wu2013Arrival,Hric2017Stochastic}. In some systems, such contact networks of customers, the patterns of nodes' leaving and entering the system can even be of the main interest \cite{Dasgupta2008Social,Kawale2009Churn,Saramaki2015Seconds}. However, there is a relatively few methods for analysing networks where both nodes and links change, and the temporal investor networks introduced here could serve as a good example for network analysis in the future research.


Additionally, in the present paper, the set of investors were based on the status of household, financial institution, or non-financial institution and activeness, which is rather arbitrary way to classify investors. Also, one could say that the observations of investor trading events are just realizations of a non-observable (psychological) process, making the identified temporal network unstable. In out future research, we will develop sampling methods to overcome potential these problems. Also, alternative inference techniques for the estimation of network edges are expected in the future research. 

\section*{Materials and methods}


\subsection*{Data}

The data used in this study is the central register of shareholdings for Finnish stocks from Finnish central depository, provided by Euroclear Finland. It includes all the major publicly traded Finnish stocks from 1995. It consists of shareholdings of all the Finnish and non-Finnish investors traded in the Helsinki stock exchange on a daily level basis. The data contains investors' trades and portfolios including all Finnish household investors, Finnish institutions, and foreign institutions. The records are exact duplicates of the official certificates of ownership and trades, and hence are very reliable. The Book Entry System entails compulsory registration of holdings for Finnish individuals (referred to as households) and institutions. Foreigners are partially exempt from registration as they can opt for registration in a nominee name, and thus they cannot be separated from each other, for which reason data about foreigners trades is excluded in the present paper. A more detailed descriptions of the data set is provide in Refs. \citen{grinblatt2000investment,tumminello2012identification}. 

Our sample data consists of marketplace transactions of  \textbf{Nokia} stock consisting of investors transactions from 1 January 1998 to December 2002. Each data record has following information: stock ticker, owner id, trading date, transaction registration date, number of shares traded, the price of trade, buy/sell transaction type, and other investor specific fields like investors' sector code, language code, gender, date of birth, and postal code. We have considered investors from different categories who have traded actively with Nokia for our analysis.

\subsection*{Links in the network}

Net volume traded by an investor $i$ on day $t$ is given as $V_{i,t}=V^b_{i,t} - V^s_{i,t}$, here \( V^b_{i,t} \) is the number of shares of Nokia bought by investor $i$ on day $t$ and \( V^s_{i,t} \) is the number of shares of Nokia sold by investor $i$ on day $t$. 
In comparison to the inference method introduced in ref. \citen{tumminello2012identification}, we do not scale the net-volumes by $V^b_{i,t} + V^s_{i,t}$, because the scaled approach does not measure the magnitude of trades, i.e. the level of the scaled variable does not reflect exceptionally high or low traded net volumes. For example, suppose that on a given day for a given stock, investor A buys one share and sells zero and investor B buys exceptionally many shares, say 1,000,000 and sells zero. Then both investors' scaled net-volumes would equal $+1$, although their trading behavior have been very different.
The dependency between two investors, $i$ and $j$, is measured with Pearson correlation for $M$ different time windows of fixed width $W$. In our study, $W$ is set to 126 trading days (6 months) and the analysis is ran with a rolling window of 1 month (21 trading days). 
As the total number of days in our data is $1252$ these choices give us $M=54$ time windows for the 6 month time window.
Note the data studied here is very sparse in a sense that for many investors most of days are without any activity (see plot (a) in Fig. \ref{fig:no_of_nodes_2000}), but these silent days are here considered as decisions for not to trade. That is, the inactive days are not considered as missing data in our calculation of the Pearson correlation coefficient.
In our notation, $\rho_t^{(ij)}$ denotes the Pearson correlation coefficient between investors $i$ and $j$ estimated from daily net-volumes of $W$ days counted
backwards from the day $t$. One could also use daily net-volumes of W/2 days in past and W/2 days in the future, but we prefer to use the data in the past instead of using the data in the future in order to analyze pre-reactions in the networks so that no information about the forthcoming bubble burst is not used.

The average absolute change in correlations between nodes that remain in two consecutive
time windows is defined as 
\begin{equation}
J_{edges}(t) = \frac{1}{|\textbf{e}_{t} \cap \textbf{e}_{t+1}|} \sum_{(i,j) \in  \textbf{e}_{t} \cap \textbf{e}_{t+1}} \left(|{\rho_{t+1}^{(ij)} - \rho_{t}^{(ij)}|}\right) \,,
\label{eq:jaccard_edges}
\end{equation}
where $\mathbf{e}_t$ denotes the set of edges in the network at time $t$ (i.e., $\mathbf{e}_t = \{ (u,v) \, |\, u,v \in \mathbf{n}_t,\, u\neq v\} $).

\subsection*{Minumum and maximum spanning trees}

For a network with $N_t$ nodes and edge set $\mathcal{E}_t$, a maximum spanning tree is a connected sub-network with the same nodes and a subset of $N_t-1$ edges $\mathcal{E}^{max}_t \subseteq \mathcal{E}_t$ such that the sum of the edge weights (here correlations), $\sum_{{(i,j)}\in \mathcal{E}_t^{max}} {\rho_t^{(ij)}}$, is maximized. Similarly, for a minimal spanning tree we find a set of edges $\mathcal{E}^{min}_t$ such that the sum of the edge weights is minimized. 

Note that we do not transform the correlations into distance using formula $d_{ij}= \sqrt{2(1-\rho_t)}$, which would make minimal spanning trees to maximal ones and vice-versa -- spanning tree structure is otherwise invariant to this transformation because this transformation only reverses the rank-order of the edge weights. We also construct minimum spanning trees, which are complementary to the maximum ones.

The average weights of maximum and minimum spanning trees are defined as:
\begin{equation}\nonumber
 L_{max}(t) =\frac{1}{(N_t-1)} \sum_{(i, j) \in {\mathcal{E}_t^{max}}}{\rho_t^{(ij)}}.
\end{equation}
and
\begin{equation}\nonumber
 L_{min}(t) =\frac{1}{(N_t-1)} \sum_{(i, j) \in {\mathcal{E}_t^{min}}}{\rho_t^{(ij)}},
\end{equation}
respectively.


%
%
%

\bibliography{references}

\end{document}